# Time-Reversal-Invariant Altermagnetic Acoustic Crystals

Tianzhi Xia,[1,2*] Han-Rong Xia,[1,*,†] Jinglin Liu,[1,*] Xiying Fan,[3] Zebin Zhu,[1] and Zhen Gao[1,§]

[1] State Key Laboratory of Optical Fiber and Cable Manufacturing Technology, Department of Electronic and Electrical Engineering, Guangdong Key Laboratory of Integrated Optoelectronics Intellisense, Southern University of Science and Technology, Shenzhen 518055, China

[2] College of Physics and Electronic Engineering, Qujing Normal University, Qujing 655011, China

[3] Department of Physics, Hubei University, Wuhan 430062, China

*These authors contributed equally to this work.

Corresponding Email: xiahr@sustech.edu.cn, gaoz@sustech.edu.cn

**Abstract:** Altermagnets have emerged as a new class of magnetic materials that combine spin-split electronic bands with zero net magnetization. Extending this paradigm to classical-wave systems has, however, been fundamentally challenging because conventional realizations require broken time-reversal symmetry (TRS). Here, we overcome this limitation by introducing a pseudospin degree of freedom and constructing a pseudo-time-reversal operator that faithfully reproduces the action of its physical counterpart while preserving actual TRS. Building on this framework, we theoretically propose and experimentally realize the first time-reversal-invariant altermagnetic acoustic crystal. Acoustic measurements directly reveal pseudospin-dependent band splitting—a defining hallmark of altermagnetism—under strictly TRS-preserving conditions. Moreover, the altermagnetic acoustic crystal exhibits sublattice–pseudospin locking, enabling flexible control over acoustic pseudospin splitting and filtering. Our work establishes acoustic crystals as a versatile platform for exploring altermagnetic physics and opens new avenues for spin-inspired wave manipulation in nonmagnetic devices.

**Introduction:** Altermagnetism, a novel form of magnetism beyond traditional ferromagnetism and antiferromagnetism, has recently attracted considerable attention in both fundamental physics and spintronic device applications [1–7]. Altermagnets combine the characteristics of both ferromagnets and antiferromagnets: they exhibit spin-split bands akin to ferromagnets, yet possess vanishing net magnetization as in antiferromagnets [1–3]. This unique combination arises from a distinct magnetic symmetry that allows time-reversal symmetry (TRS) breaking without a global magnetic moment [8], leading to momentum-dependent spin polarization and novel transport phenomena, such as anomalous Hall effect [9–12] and spin-to-charge conversion [13]. The experimental confirmation of altermagnetism in electronic materials such as MnTe and CrSb has established this phase as a promising platform for next-generation spintronic and orbitronic devices [14–20].

The appeal of altermagnetism, which offers spin splitting without net magnetization, has naturally motivated efforts to transplant its physics and applications into classical wave systems [21-24]. However, a fundamental obstacle stands in the way: the established framework of altermagnetism inherently relies on broken TRS [2,8,17,25–33]. The characteristic spin-split bands and momentum-dependent spin polarization emerge only when TRS is absent, a condition that is naturally fulfilled in magnetic materials [34–40] but is exceptionally difficult to meet in most classical wave systems [41–48]. Recently, this obstacle has been circumvented in photonics along two routes: by genuinely breaking TRS using magneto-optical materials [22,23], or, while preserving TRS, by exploiting the helicity of light in chiral structures—a bosonic degree of freedom that is even under time reversal [21,24]. Neither route is readily available in acoustics. Conventional approaches to breaking TRS, such as fluid-flow biasing [49] or active components [50], introduce substantial complexity and loss and are difficult to reconcile with on-chip integration. Moreover, airborne sound is a scalar wave and therefore possesses no intrinsic polarization or helicity. This raises a fundamental question at the heart of classical-wave altermagnetism: can its defining hallmarks—momentum-dependent spin splitting and sublattice-locked spin textures—emerge in a system that strictly preserves TRS and lacks any helicity degree of freedom?

Here, we answer this question affirmatively by introducing a pseudospin degree of freedom on a bilayer centered square lattice with layer-dependent potentials and staggered interlayer couplings, all within a perfectly time-reversal-invariant framework. We construct a pseudo-time-reversal operator that connects the two pseudospin sectors by flipping the pseudospin while preserving true TRS. By tuning the next-nearest-neighbor (NNN) hoppings, we control the effective interactions among identical pseudospin species along the $x$ and $y$ directions. Isotropic interactions produce doubly degenerate bands, corresponding to an antiferromagnetic phase, whereas anisotropic interactions—characterized by distinct $x$ and $y$ hoppings for the same pseudospin—break the pseudo-TRS and induce momentum-dependent pseudospin splitting with sublattice-spin locking, thus giving rise to an altermagnetic phase. We experimentally realize this phase in a time-reversal-invariant acoustic crystal and directly observe the characteristic spin-split bands, anisotropic iso-frequency contours, and spin-sublattice locking and filtering. A reference structure with restored pseudo-TRS shows no such splitting, confirming its altermagnetic origin. Our approach requires no genuine TRS breaking and can be directly extended to photonics, mechanics, and other classical wave systems, establishing a general framework for simulating magnetic symmetries in nonmagnetic, time-reversal-invariant systems.

**Tight-binding model:** We consider a bilayer centered square lattice with lattice constant $a$, whose two sublattices $A$ and $B$ are offset by the vector $(a/2, a/2)$ [Fig. 1(a)]. Each site carries a layer degree of freedom (layers 1 and 2, corresponding to the lower and upper physical layers, respectively), and the two sublattices host opposite exchange fields acting on it: on sublattice $A$, the on-site potentials are $(V, -V)$ for layers 1 and 2 with interlayer hopping $\kappa$ [red lines in Fig. 1(a)]; on sublattice $B$, they are $(-V, V)$ with interlayer hopping $-\kappa$ (blue lines). In the layer basis, the two on-site (dimer) Hamiltonians are

$$H_A = \begin{pmatrix} V & \kappa \\ \kappa & -V \end{pmatrix}, H_B = \begin{pmatrix} -V & -\kappa \\ -\kappa & V \end{pmatrix}. \quad (1)$$

They are connected by a pseudo-time-reversal operator $\mathcal{T}_p = \sigma_y K$ ($\mathcal{T}_p^2 = -1$), i.e., $H_B = \mathcal{T}_p H_A \mathcal{T}_p^{-1}$. Here, $\sigma_y$ is the Pauli $y$-matrix acting on the layer space, and $K$ is the complex conjugation operator. Note that the pseudo-time-reversal operator $\mathcal{T}_p$ and the genuine fermionic time-reversal operator $\boldsymbol{T}_f$ differ only by a phase factor: $\mathcal{T}_p = -i\boldsymbol{T}_f$. Figure 1(a) labels the two exchange-field configurations as "Exchange field A" and "Exchange field B." They act as the two magnetic sublattices of an altermagnet, while physical TRS is strictly preserved. The pseudospin follows from the eigenstates of the exchange field. Using Pauli matrices $\tau_i$ and $\sigma_i$ for the sublattice and layer spaces, respectively, we define the pseudospin operator in the full $4 \times 4$ space as

$$S_z = \tau_0 \otimes \frac{V\sigma_z + \kappa\sigma_x}{\sqrt{V^2 + \kappa^2}}, \tag{2}$$

where $\tau_0$ is the identity matrix in the sublattice space. The full-space pseudo-time-reversal operator $\mathcal{T}_p = (\tau_0 \otimes \sigma_y)K$ satisfies $\mathcal{T}_p{}^2 = -I_4$ ($I_4$ being the $4 \times 4$ identity matrix) and $\mathcal{T}_p S_z \mathcal{T}_p{}^{-1} = -S_z$, confirming that $S_z$ behaves as a genuine pseudospin that is flipped by $\mathcal{T}_p$. Throughout this work, the $S_z = +1$ and $S_z = -1$ sectors are termed the positive and negative pseudospins, respectively. Additionally, in the limit of $\kappa \ll V$, $S_z$ reduces to $\tau_0 \otimes \sigma_z$, which naturally implies that pseudospin polarization is determined solely by the layer index: positive-pseudospin states reside predominantly on layer 1, and negative-pseudospin states on layer 2, independent of the sublattice.

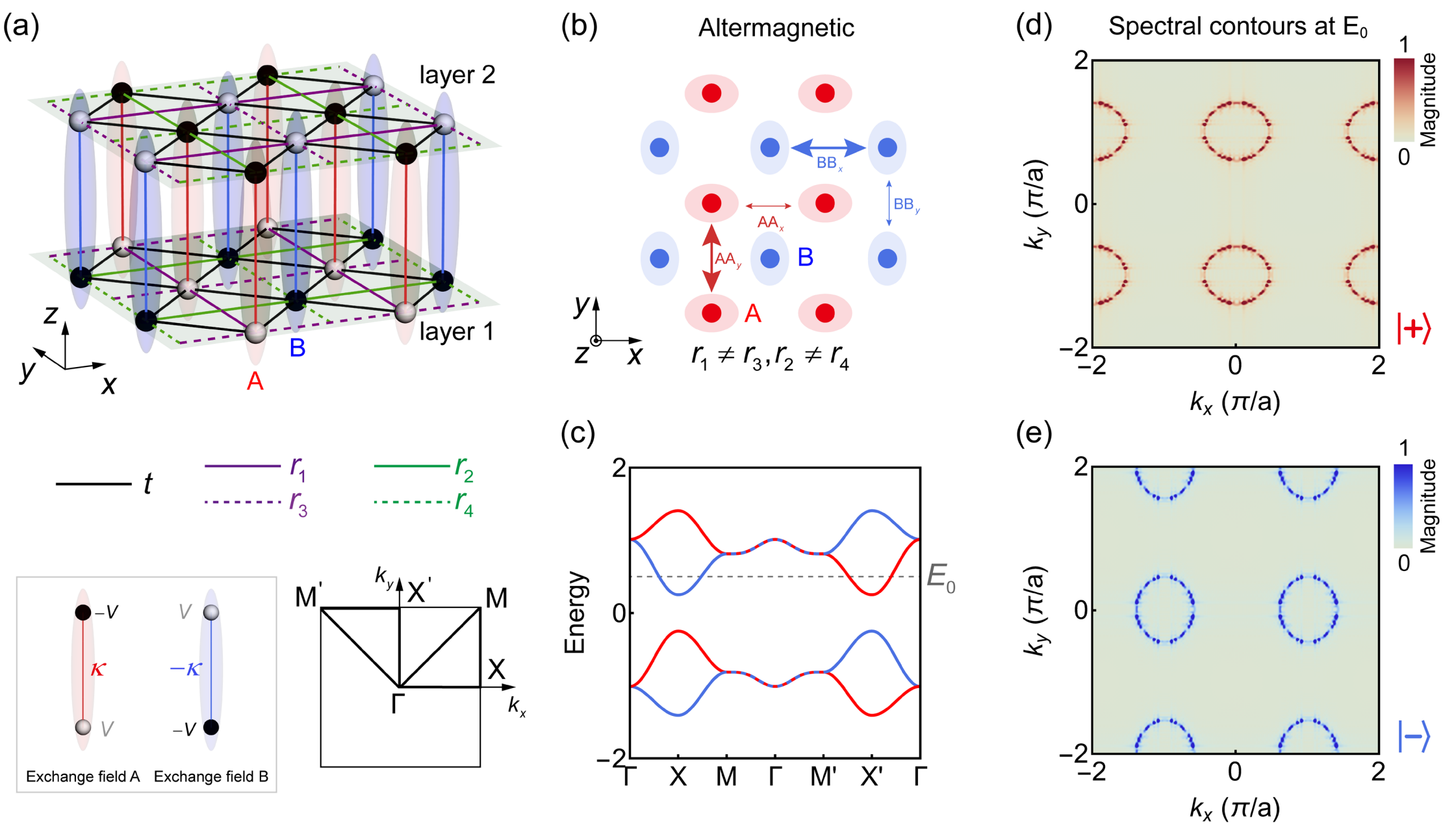

FIG. 1. Tight-binding model. (a) Top panel: a bilayer centered square lattice with two sublattices (A and B) hosting opposite exchange fields; the two horizontal planes mark layers 1 and 2. The NN hopping $t$ is fixed, while NNN couplings $r_i$ can be tuned to generate altermagnetic order. Bottom left: the two dimer configurations ("Exchange field A" and "Exchange field B") with on-site potentials $(V, -V)$ and $(-V, V)$ on layers 1 and 2 and interlayer couplings $\pm\kappa$. Bottom right: Brillouin zone with high-symmetry paths. (b) Altermagnetic phase realized under the condition $r_1 \neq r_3$ and $r_2 \neq r_4$. (c) Calculated band structure of the altermagnetic phase, revealing clear

pseudospin splitting along the $k_x$ and $k_y$ directions. (d) and (e) Calculated spectral contours at $E_0$, corresponding to the gray dashed line in (c), for pseudospin-$|+\rangle$ (d) and pseudospin-$|-\rangle$ (e), exhibiting a $C_2$ symmetry and different centering (positive pseudospin along $k_y$, negative pseudospin along $k_x$).

As shown in the top panel of Fig. 1(a), opposite exchange-field configurations separated by displacement vector $(a/2, a/2)$ are coupled via the nearest-neighbor (NN) hopping $t$ (black line), while identical configurations on sublattices separated by a lattice vector are connected via NNN couplings $r_i$ $(i = 1, 2, 3, 4)$. To write the Bloch Hamiltonian in momentum space, we adopt the basis$\{|\mathrm{A},1\rangle, |\mathrm{A},2\rangle, |\mathrm{B},1\rangle, |\mathrm{B},2\rangle\}$, where 1 and 2 label the two layers. In this basis, the Hamiltonian takes the compact form:

$$H(\boldsymbol{k}) = \epsilon I_4 + m\tau_z \otimes \sigma_z + \Delta_\tau \tau_z \otimes \sigma_0 + \Delta_\sigma \tau_0 \otimes \sigma_z + f(\boldsymbol{k})\tau_x \otimes \sigma_0 + \kappa\tau_z \otimes \sigma_x, \quad (3)$$

with

$$\epsilon = \frac{r_1 + r_2 + r_3 + r_4}{2}(\cos k_x a + \cos k_y a), \quad (4)$$

$$m = V + \frac{r_1 - r_2 + r_3 - r_4}{2}(\cos k_x a + \cos k_y a), \quad (5)$$

$$\Delta_\tau = \frac{r_1 + r_2 - r_3 - r_4}{2}(\cos k_y a - \cos k_x a), \quad (6)$$

$$\Delta_\sigma = \frac{r_1 - r_2 - r_3 + r_4}{2}(\cos k_y a - \cos k_x a), \quad (7)$$

$$f(\boldsymbol{k}) = 2t\left[\cos\left(\frac{k_x a + k_y a}{2}\right) + \cos\left(\frac{k_x a - k_y a}{2}\right)\right]. \quad (8)$$

The two terms $\Delta_\tau$ and $\Delta_\sigma$, both proportional to the common factor $(\cos k_y a - \cos k_x a)$, explicitly reveal the $d_{x^2-y^2}$-wave anisotropy of the altermagnetic phase. Crucially, the pseudospin is not merely a label but a quantum number controlled by symmetry. Direct evaluation of Eqs. (2) and (3) yields

$$[S_z, H(\boldsymbol{k})] = \frac{2i\kappa}{\sqrt{V^2 + \kappa^2}}\left[\left(V - m(\boldsymbol{k})\right)\tau_z - \Delta_\sigma(\boldsymbol{k})\right] \otimes \sigma_y, \quad (9)$$

which vanishes identically at every $\boldsymbol{k}$ whenever $r_1 = r_2$ and $r_3 = r_4$, i.e., whenever the two layers share the same NNN dispersion within each sublattice. This condition is exactly satisfied by our experimental structure (Figs. 2–4), so the pseudospin $S_z$ is strictly conserved there and all bands can be labeled by its eigenvalues. For the mixed-sign couplings used in Figs. 1(c)–1(e), the residual mixing is weak and quantifiable: the pseudospin purity $|\langle S_z \rangle|$ of the Bloch eigenstates averages 0.99 over the Brillouin zone for the parameters of Fig. 1(c) (Supplementary Text, Section S3; Fig. S1). Note that when $r_1 = r_3$ and $r_2 = r_4$, both $\Delta_\tau$ and $\Delta_\sigma$ vanish. In this case, from the symmetry perspective, the combined operation $\mathcal{L}_{(a/2,a/2)}\mathcal{T}_p$ (with $\mathcal{L}$ the lattice translation operator) leaves the system invariant, yielding pseudospin degeneracy at all $\boldsymbol{k}$ points, corresponding to an antiferromagnetic phase (see Supplementary Fig. S2(a)). When $r_1 \neq r_3$ and $r_2 \neq r_4$, as shown in Fig. 1(b), the local environment of an exchange center becomes anisotropic, such that the interactions for exchange orientations along different directions are unequal: $\mathrm{AA}_x \neq \mathrm{AA}_y$, $\mathrm{BB}_x \neq \mathrm{BB}_y$. In this case, the system cannot be restored by any lattice translation after applying the pseudo-time-reversal operator $\mathcal{T}_p$. Consequently, spin splitting occurs along the $k_x$ and $k_y$ directions in the standard centered square lattice, accompanied by the selective arrangement of positive and negative pseudospins onto the sublattices, namely an altermagnetic state. (See Supplementary Fig. S3 for the pseudospin splitting versus the NNN coupling ratios.) The residual degeneracy along the diagonal lines $k_x = \pm k_y$ (i.e., Γ–M and Γ–M′), where $\Delta_\tau$ and $\Delta_\sigma$ vanish, is likewise symmetry-protected: the above combined operation acts on these lines as the antiunitary $\Theta = (\tau_x \otimes \sigma_y)K$, which satisfies $\Theta^2 = -I_4$ and commutes with $H(\boldsymbol{k})$ pointwise, protecting the Kramers-like nodal degeneracy characteristic of d-wave altermagnetism (Supplementary Text, section S2).

Having established the system Hamiltonian and the resulting altermagnetic phase, we then focus on the spin splitting in momentum space. Figure 1(c) shows the band structure of the altermagnetic phase with parameters $\{V, \kappa, t, r_1, r_2, r_3, r_4\} = \{0.8, 0.15, -0.15,\ 0.15, -0.15, -0.15, 0.15\}$ (See Supplementary Fig. S2 for band structures with other typical parameters). We further perform Green's function calculations to excite the field at energy $E_0$: $\psi = [(E_0 + i\gamma)I - H]^{-1} \cdot \psi_{source}$, where the loss is accounted for by the imaginary term $i\gamma$. The constant-energy contours (analogous to Fermi surfaces) at $E_0$ are then obtained via Fourier transform of the field. As shown in Fig. 1(d) and Fig. 1(e), the constant-energy contours for positive (red) and negative (blue) pseudospin reduce to $C_2$ symmetry about the origin in momentum space (evaluated with $E_0 = 0.5, \gamma = 0.001$, and a system size of $75 \times 75$ unit cells). Moreover, the positive-pseudospin contours are centered on the $k_y$-axis ($k_x = 0$), while the negative-pseudospin contours are centered on the $k_x$-axis ($k_y = 0$), both exhibiting an anisotropic signature. The reduction from $C_4$ to $C_2$ symmetry and the opposite centering of two pseudospin contours reflect the underlying sublattice-spin locking induced by broken pseudo-TRS $\mathcal{T}_p$, marking a key signature of the altermagnetic phase in momentum space.

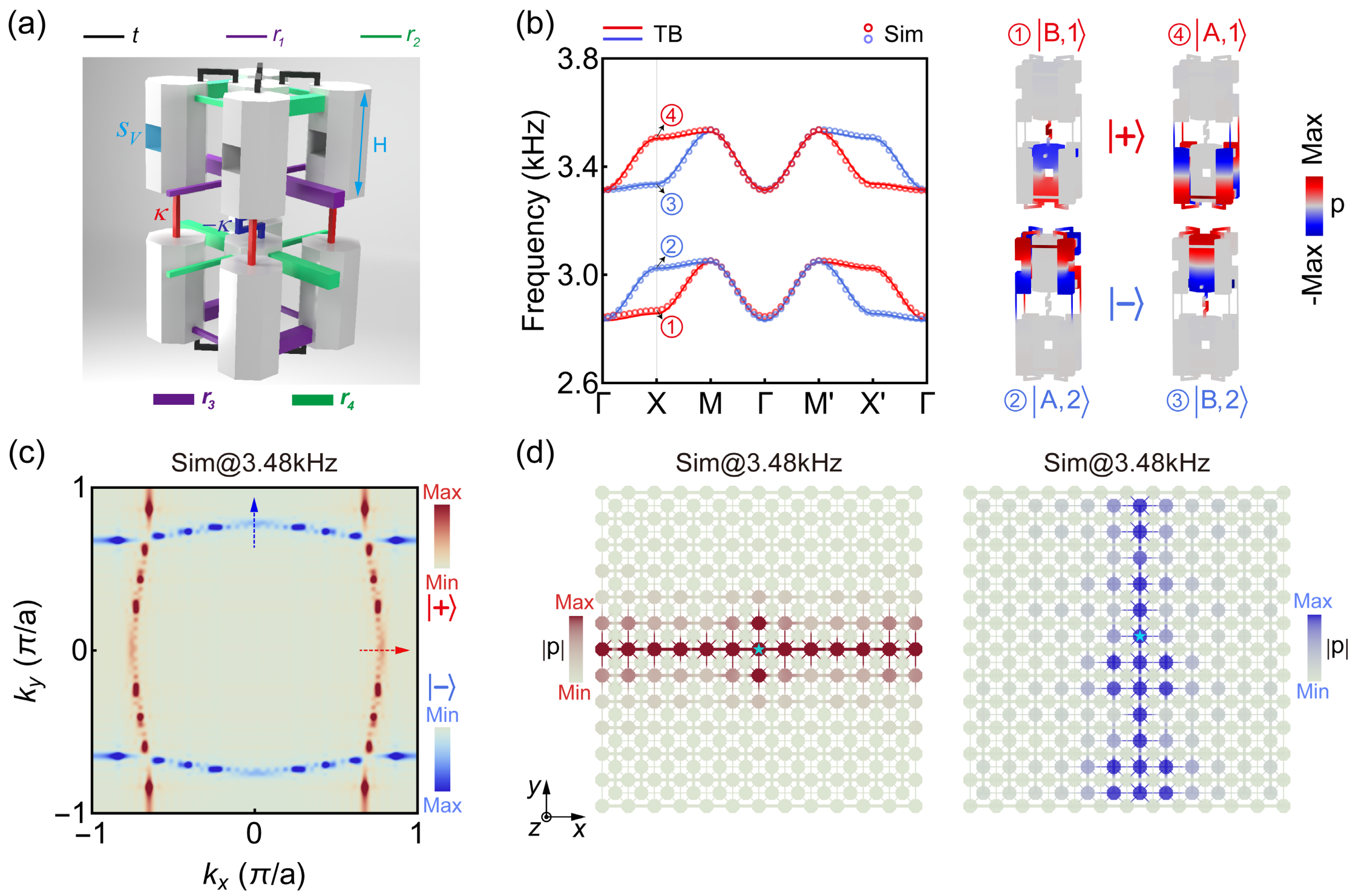

FIG. 2. Acoustic realization of the altermagnetic phase. (a) Acoustic cavity-tube structure for realizing the altermagnetic phase. Cavities without and with a central hole correspond to on-site potentials $+V$ and $-V$ (the hole lowers the on-site potential by $2V$), respectively. The average frequency of the upper and lower cavities determines the center frequency. The coupling strengths are controlled by the cross-sectional areas of the connecting tubes. (b) Left panel: eigenfrequency spectra along the high-symmetry line of the Brillouin zone. The tight-binding results are shown as solid lines, calculated with $\omega_0 = 3.18$ kHz, $V = 0.24$ kHz, $t = -\kappa = r_1 = r_2 = -0.006$ kHz, and $r_3 = 8r_1 = r_4 = 8r_2$, while the COMSOL simulation results are represented by circles. Red and blue denote the positive and negative pseudospin states ($S_z = \pm 1$), respectively. Right panel: four eigenfield profiles at the $X$ point (labeled by numbers in ascending frequency order). Their (sublattice, layer) contents are predominantly $|\mathrm{B},1\rangle$, $|\mathrm{A},2\rangle$, $|\mathrm{B},2\rangle$, and $|\mathrm{A},1\rangle$: with the tight-binding basis ordered from layer 1 to layer 2, the positive-pseudospin modes ① and ④ ($|+\rangle$) are polarized on layer 1, and the negative-pseudospin modes ② and ③ ($|-\rangle$) on layer 2. (c) Simulated iso-frequency contours of the altermagnetic acoustic crystal at 3.48 kHz. For both positive (dark red) and negative (dark blue) pseudospin states, two slightly curved contours appear along the $k_y$ and $k_x$ directions, respectively. The anisotropic shape of these contours directly reflects the pseudospin-dependent band splitting and indicates orthogonal propagation directions: positive-pseudospin states propagate predominantly along the $x$ direction (with rapid decay along $y$), whereas negative-pseudospin states propagate predominantly along the $y$ direction (with rapid decay along $x$). (d) Simulated acoustic field distributions at 3.48 kHz demonstrating pseudospin–sublattice locking. Under positive-pseudospin excitation (left panel), the acoustic field is localized on sublattice A and propagates along the $x$ direction, while decaying along the $y$ direction. Under negative-pseudospin excitation (right panel), the field is localized on sublattice B and propagates along the $y$ direction, while decaying along the $x$ direction, consistent with the iso-frequency contours in (c).

**Acoustic realization of the altermagnetic phase:** We now realize the aforementioned tight-binding model in an acoustic crystal, in which the orbitals and hoppings are emulated by acoustic cavities and narrow tubes. Previous studies have demonstrated that the on-site potential can be linearly tuned via the cross-sectional area of a hole drilled at the cavity center [51], while the coupling strength can be linearly controlled by the cross-sectional area of the connecting tubes [52–56], thus facilitating a straightforward implementation of our theoretical model. As illustrated in Fig. 2(a), the altermagnetic acoustic crystal comprises two layers of cavities—each with a cross-sectional circumradius $\rho = 15$ mm, a height $H = 50$ mm, and a lattice constant $a = 52.72$ mm, arranged on a centered square lattice. The on-site potentials $+V$ and $-V$ are realized by acoustic cavities without and with a central hole of area $S_V = 84.73$ mm², respectively. The coupling tubes for the NN ($t$), interlayer ($\kappa$), and NNN ($r_1$ and $r_2$) hoppings are designed to share an identical cross-sectional area of $S_i = 4.69$ mm². In contrast, the remaining NNN couplings $r_3$ and $r_4$ are implemented using air tubes with a larger cross-sectional area, specifically $S_{r3} = S_{r4} = 37.52$ mm². Figure 2(b) presents the simulated band structure of the acoustic crystal (left panel) alongside its four eigenfield profiles at the $X$ point (right panel). These results clearly reveal pseudospin splitting and pseudospin-sublattice locking, both hallmarks of the altermagnetic phase. In particular, the four X-point modes acquire, in ascending frequency order, the (sublattice, layer) contents $|\mathrm{B},1\rangle$, $|\mathrm{A},2\rangle$, $|\mathrm{B},2\rangle$, and $|\mathrm{A},1\rangle$ (the exact eigenstates are weakly rotated layer combinations, see Supplementary Text, section S3). Since our basis is ordered from layer 1 to layer 2, the positive-pseudospin modes ① and ④ are polarized on layer 1, and the negative-pseudospin modes ② and ③ on layer 2, in full agreement with the tight-binding analysis. Note that the acoustic band structure differs slightly from the tight-binding prediction owing to experimental feasibility constraints. Supplementary Fig. S4 presents an alternative parameter set that yields a band dispersion more closely resembling the tight-binding result.

We now examine the distinctive features of the anisotropic iso-frequency contours. Figure 2(c) shows the iso-frequency contours computed via the Green's function method on a $75 \times 75$ supercell, using parameters extracted by fitting the tight-binding Hamiltonian to the band structure in Fig. 2(b). For the positive (dark red) and negative (dark blue) pseudospin states, the contours consist of two slightly curved lines along the $k_y$ and $k_x$ direction, respectively. These anisotropic iso-frequency contours directly reflect the pseudospin-dependent band splitting and indicate opposite directional propagation: positive-pseudospin states propagate predominantly along the $x$ direction (with rapid decay along $y$), whereas negative-pseudospin states propagate predominantly along the $y$ direction (with rapid decay along $x$), thus enabling pseudospin filtering. To directly visualize this pseudospin–sublattice locking, we perform point-source excitation simulations at 3.48 kHz. As shown in Fig. 2(d), when the source (cyan star placed on sublattice A) excites the positive-pseudospin states (left panel), the field is localized on sublattice A and propagates along the $x$ direction while decaying along $y$ direction. Conversely, when the source (cyan star placed on sublattice B) excites the negative-pseudospin states (right panel), the field localizes on sublattice B and propagates along the $y$ direction while decaying along $x$ direction. These propagation directions are consistent with the iso-frequency contours in Fig. 2(c), confirming the unique pseudospin–sublattice locking feature of the altermagnetic acoustic crystal. In the finite-element simulations, acoustic dissipation was considered by using a complex sound velocity $c = (343 + 0.5i)$ m/s, to match the experimental conditions. Notably, the eigenfield profiles in Fig. 2(b) also reveal layer-selective pseudospin localization. At 3.48 kHz, excitation on the A sublattice (positive-pseudospin ④) concentrates energy predominantly on the layer 1, whereas excitation on the B sublattice (negative-pseudospin ③) localizes it on the layer 2, with the remaining layer having almost no energy distribution. (Note that only the energy-carrying layer is shown in the acoustic field distributions [Fig. 2(d)]. The complete layer-resolved field profiles in Supplementary Fig. S5, where both layers are driven simultaneously and the complementary layer shows essentially no propagation, further justify this single-layer excitation protocol.)

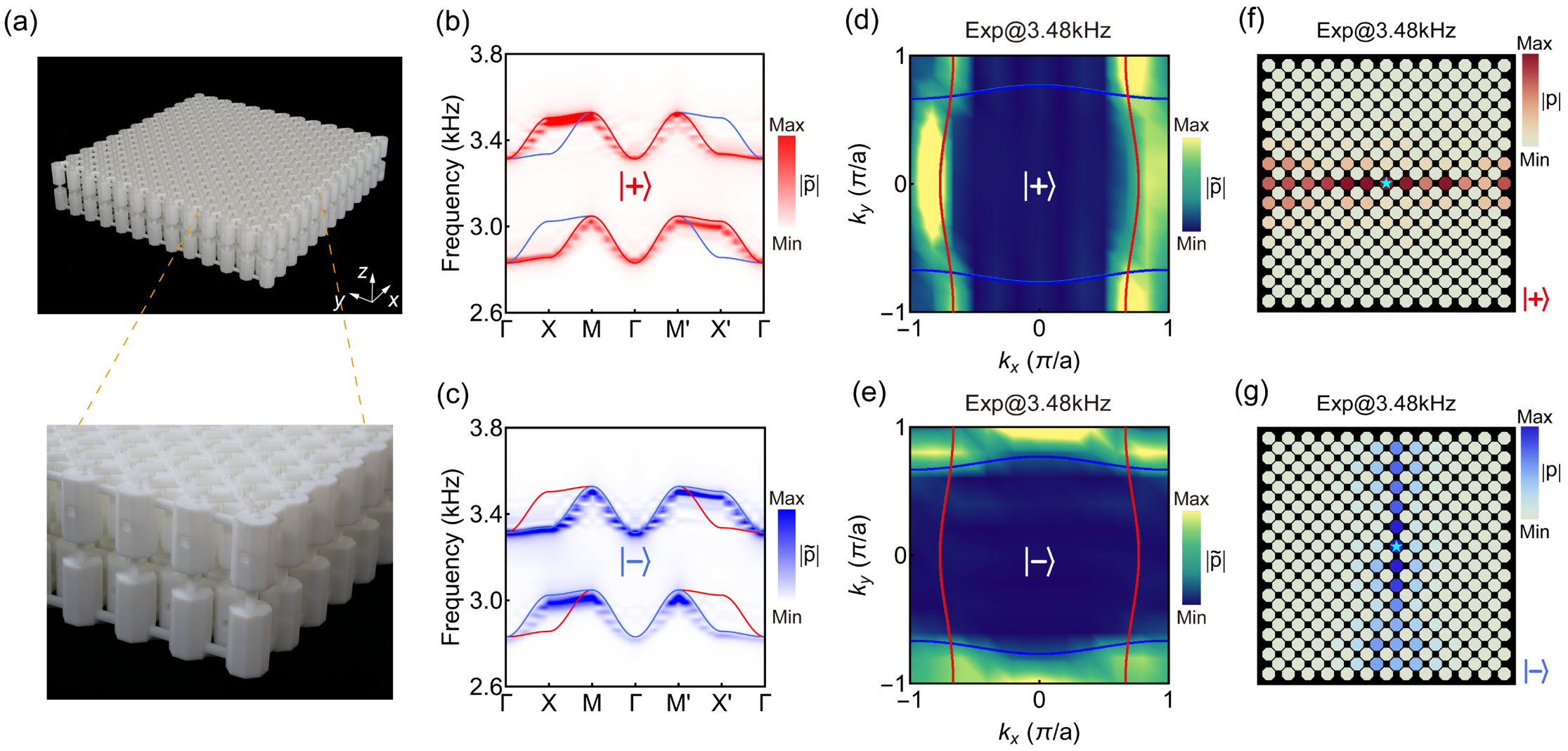

FIG. 3. Experimental demonstration of the altermagnetic acoustic crystal. (a) Photograph of the fabricated acoustic sample. (b-c) Measured (colormap) and theoretical (red and blue solid lines) bulk band structures along the high-symmetry momentum path under layer-1 and layer-2 excitation, which select the positive ($|+\rangle$) and negative ($|-\rangle$) pseudospin sectors, respectively. The bands display a pronounced spin splitting. (d-e) Measured (colormap) and theoretical (red and blue solid lines) iso-frequency contours at 3.48 kHz for the positive (d) and negative (e) pseudospin states. The iso-frequency contours are elongated in mutually orthogonal directions (vertical versus horizontal), thereby directly revealing the anisotropic band dispersion and confirming pseudospin-momentum locking in the altermagnetic phase. (f-g) Measured acoustic pressure distributions at 3.48 kHz for positive-pseudospin (f) and negative-pseudospin (g) excitations at different sublattices A and B, respectively, exhibiting field confinement to different sublattices and orthogonal propagation directions.

**Experimental observation of the acoustic altermagnetic phase:** To demonstrate the acoustic altermagnetic phase, we fabricated a bilayer-centered square lattice acoustic crystal using the design parameters described above. Figure 3(a) shows a photograph of the sample, which was 3D-printed from photosensitive resin with a fabrication tolerance of approximately 0.1 mm (See Supplementary Fig. S6 for the top view of the sample and unit-cell marking). To excite and detect acoustic waves inside the cavities, small holes were drilled into the resonators to accommodate a sound source or a probe; the holes were sealed when not in use. In our experiments, we placed a pair of acoustic sources at the center of the sample and scanned the sound-pressure field cavity by cavity. Both the input and output signals were recorded and frequency-resolved using a vector network analyzer (Keysight E5061B). By performing a two-dimensional spatial Fourier transform, we obtained the spin-resolved bulk band structure in momentum space. Because the pseudospin eigenstates are layer-polarized, layer-selective excitation directly addresses the two pseudospin sectors. Figures 3(b) and 3(c) present the measured bulk band structures (colormap) under layer-1 and layer-2 excitation, i.e., for the positive and negative pseudospin sectors, respectively, in excellent agreement with the theoretical bands (red and blue lines). Notably, both the measured and simulated bulk band structures exhibit a clear splitting between the two pseudospin states, a hallmark of the altermagnetic phase. We further extracted the iso-frequency contours at 3.48 kHz from the momentum-space Fourier data. As shown in Figs. 3(d) and 3(e), the measured (colormap) and simulated (red and blue lines) positive- and negative-pseudospin iso-frequency contours are elongated in orthogonal directions—vertical and horizontal, respectively—which directly reflect the anisotropic band dispersion. These orthogonal anisotropic iso-frequency contours provide strong evidence of pseudospin-momentum locking, another characteristic signature of the altermagnetic phase in our acoustic system. To directly verify pseudospin–sublattice locking in real space, we performed selective excitations on individual sublattices and measured the resulting acoustic pressure field distributions. Figures 3(f) and 3(g) display the measured acoustic field distributions at 3.48 kHz for positive-pseudospin excitation (cyan star) on sublattice A and negative-pseudospin excitation (cyan star) on sublattice B, respectively. In Fig. 3(f), the acoustic field is predominantly confined to sublattice A and propagates along the $x$ direction, whereas in Fig. 3(g), the field is mainly localized on sublattice B and propagates along the $y$ direction (See Supplementary Fig. S7 for the frequency-resolved intensities under all four (sublattice, layer) excitation channels), consistent with the anisotropic iso-frequency contours in Figs. 3(d) and 3(e). To unambiguously establish that the observed splitting is altermagnetic in origin, we fabricated and measured a reference structure [Fig. S8] in which the NNN couplings are pairwise equal ($r_1 = r_3$, $r_2 = r_4$), so that the combined translation–pseudo-TRS $\mathcal{L}(a/2, a/2)\mathcal{T}_p$ is restored (antiferromagnetic phase). As shown in Supplementary Fig. S8, both the simulated and measured bands of this reference structure remain doubly degenerate along the entire high-symmetry path, and the two pseudospin-resolved spectra are identical within experimental uncertainty—no pseudospin splitting is observed. This control experiment confirms that the splitting, the anisotropic contours, and the directional filtering reported above arise from the broken pseudo-TRS of the altermagnetic phase.

Finally, we explore a key functionality arising from the pseudospin degrees of freedom—pseudospin filtering. As schematically illustrated in Fig. 4(a), when both positive- and negative-pseudospin states are simultaneously excited inside the altermagnetic structure, the two channels propagate in orthogonal directions, enabling the extraction of a single pseudospin species at each output port. We first verify this concept via numerical simulations. Figure 4(b) shows the simulated acoustic pressure field distribution at 3.48 kHz, where four sources (stars) placed at the lower-left corner excite both pseudospin channels. The positive-pseudospin states (A sublattice) travel along the $x$ direction, while the negative-pseudospin states (B sublattice) propagate along the $y$ direction, in agreement with the anisotropic iso-frequency contours in Fig. 2(c). To experimentally demonstrate this effect, we employ the source configuration shown in Fig. 4(c), in which four sources at the lower-left corner simultaneously excite both pseudospin states. As shown in Fig. 4(d), at the frequency of 3.48 kHz, the two pseudospin components separate spatially: the positive-pseudospin channel propagates along the $x$ direction, while the negative-pseudospin channel propagates along the $y$ direction, each decaying rapidly along the transverse direction. These results demonstrate that the time-reversal-invariant acoustic crystal not only reproduces the pseudospin-splitting phenomenon in altermagnets but also enables pseudospin-selective transport, paving the way for acoustic spintronic devices.

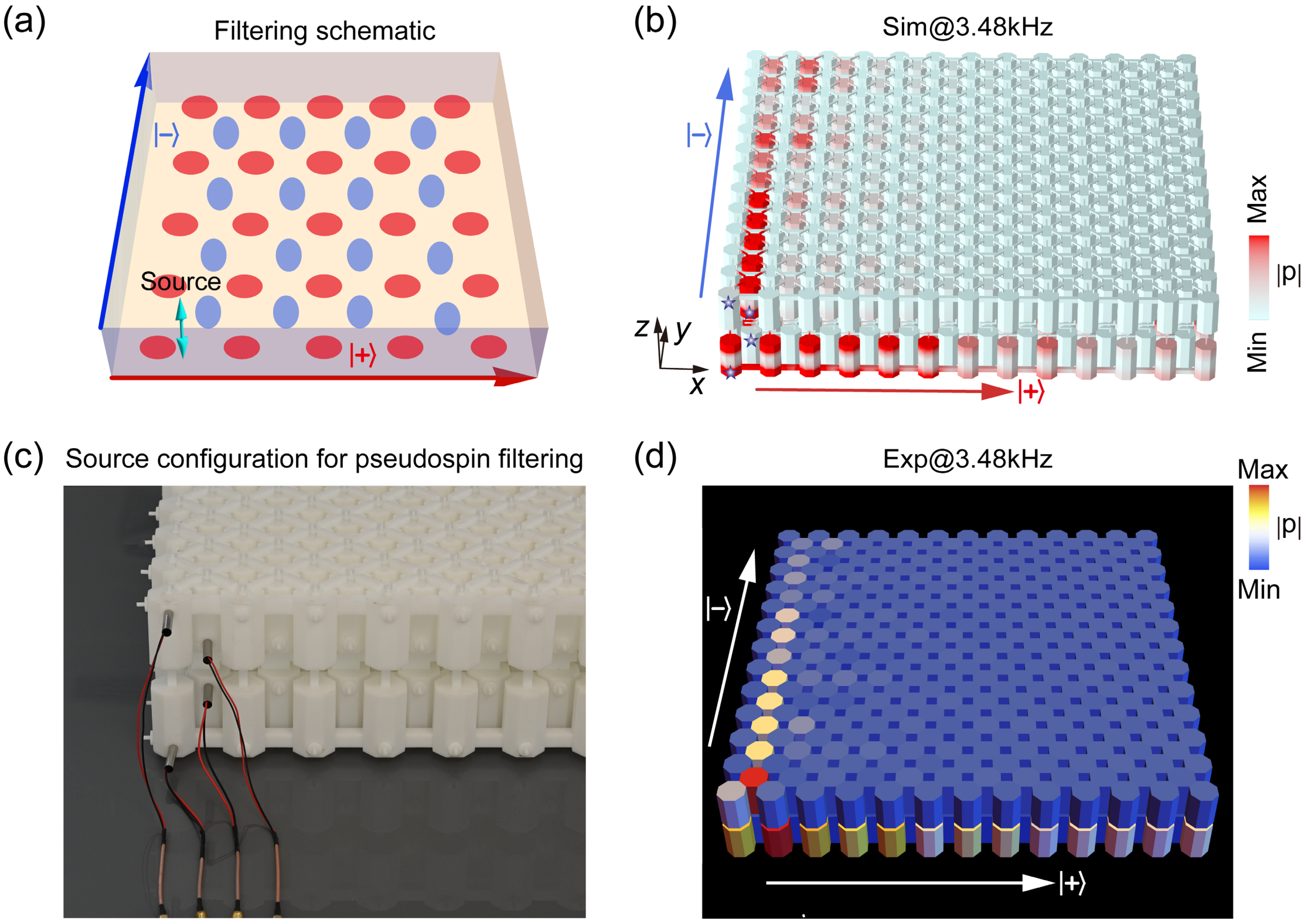

FIG. 4. Experimental characterization of acoustic pseudospin filtering. (a) Schematic illustration of pseudospin filtering in the altermagnetic structure. Upon simultaneous excitation of positive-pseudospin and negative-pseudospin states, the two channels are routed along orthogonal directions, enabling spatial separation and the selective extraction of a single pseudospin species at each output port. (b) Simulated acoustic pressure distribution demonstrating pseudospin filtering. Four sources (stars) located in the lower-left corner excite both the positive-pseudospin and negative-pseudospin states at 3.48 kHz. The positive-pseudospin states (A sublattice) propagate along the $x$ direction (red arrow), while the negative-pseudospin states (B sublattice) propagate along the $y$ direction (blue arrow). (c) Source configuration for simultaneous excitation of both positive-pseudospin and negative-pseudospin states. (d) Measured acoustic pressure distributions of the altermagnetic acoustic crystal under simultaneous excitation of both pseudospin states, the two pseudospin states propagate orthogonally along the $x$ and $y$ directions, respectively, with rapid attenuation in the transverse direction.

**Conclusion**:

In summary, we have theoretically proposed and experimentally realized a time-reversal-invariant altermagnetic acoustic crystal based on a bilayer-centered square lattice. By constructing a pseudo-time-reversal operator that faithfully mimics its genuine counterpart, we circumvent the long-standing challenge of realizing altermagnetic phases in classical-wave systems without explicitly breaking physical TRS. Although all couplings in our model are strictly real, it exhibits pseudospin-dependent band splitting, a defining hallmark of altermagnetism. Crucially, the pseudospin, defined by the eigenstates of the exchange field, is exactly conserved in the ideal tight-binding model with parameters extracted from the experimentally realized structure, so that all bands carry a well-defined pseudospin label. The nodal degeneracies along $\Gamma-\mathrm{M}$ and $\Gamma-\mathrm{M}'$ are protected by an antiunitary sublattice-exchange–spin-flip symmetry $\Theta$ with $\Theta^2=-1$. Experimentally, we directly observe both pseudospin-split bands and pseudospin–sublattice locking. The measured isofrequency contours of our sample are elongated along mutually orthogonal directions for the two pseudospin states, providing direct evidence of pseudospin–momentum locking. Furthermore, we demonstrate directional pseudospin filtering: the two pseudospin components propagate along orthogonal paths while decaying rapidly in the transverse direction. Our work establishes acoustic crystals as a versatile platform for exploring altermagnetic physics and developing acoustic spintronic devices. More broadly, this scheme is readily extendable to other classical-wave platforms, including photonic and mechanical systems, opening a route toward pseudospin-based devices that operate without breaking TRS.

**Acknowledgements**. Z.G. acknowledges funding from the National Key R&D Program of China (grant no. 2025YFA1412300), National Natural Science Foundation of China (grant no. 62375118 and 62361166627), Guangdong Basic and Applied Basic Research Foundation (grant no. 2024A1515012770), Shenzhen Science and Technology Innovation Commission (grants no. 20230802205352003), and High-level Special Funds (grant no. G03034K004).

**Data availability**. All data that support the plots within this paper and other findings of this study are available from the corresponding authors upon reasonable request.